%% file: main.tex
\documentclass[letterpaper]{article} 
\usepackage[draft]{aaai2026}  
\usepackage{times}  
\usepackage{helvet}  
\usepackage{courier}  
\usepackage[hyphens]{url}  
\usepackage{graphicx} 
\usepackage{natbib}  
\usepackage{caption} 
\usepackage{algorithm}
\usepackage{algorithmic}
\usepackage{xcolor}
\usepackage{xspace} 
\usepackage{amsfonts}
\usepackage{multirow}
\usepackage{booktabs}
\usepackage{caption}
\usepackage{makecell}
\usepackage[most]{tcolorbox}
\definecolor{pink_our}{HTML}{F7CECD}
\definecolor{blue_our}{HTML}{D6E0F3}
\definecolor{green_our}{HTML}{D9E7D6}
\definecolor{yellow_our}{HTML}{FCF1CC}
\definecolor{purple_our}{HTML}{DFD5E6}
\usepackage{subfig}
\newcommand{\eat}[1]{}
\newcommand{\addRed}[1]{\textcolor{black}{#1}}
\newcommand{\baby}{\textsc{KGTB}\xspace}

\usepackage{newfloat}
\usepackage{listings}
\DeclareCaptionStyle{ruled}{labelfont=normalfont,labelsep=colon,strut=off} 
\floatstyle{ruled}
\newfloat{listing}{tb}{lst}{}
\floatname{listing}{Listing}
\title{Knowledge Graph Tokenization for Behavior-Aware Generative Next POI Recommendation}
\author {
    Ke Sun\textsuperscript{\rm 1},
    Mayi Xu\textsuperscript{\rm 2}\thanks{Corresponding author.}
}
\affiliations {
    \textsuperscript{\rm 1}School of Computer Science and Technology, Wuhan University of Science and Technology, Wuhan, China\\
    \textsuperscript{\rm 2}School of Computer Science, Wuhan University, Wuhan, China\\
    sunke@wust.edu.cn, xumayi@whu.edu.cn
}

\usepackage{bibentry}

\begin{document}

\maketitle

\begin{abstract}
Generative paradigm, especially powered by Large Language Models (LLMs), has emerged as a new solution to the next point-of-interest (POI) recommendation. 
\eat{The main challenge is how to effectively make complex recommendation data compatible with LLMs and activate LLMs' understanding of human mobility behavior. Pioneering studies achieve this with a two-stage pipeline, starting with a tokenizer converting POIs to discrete identifiers, followed by POI behavior prediction tasks for LLM fine-tuning.}
Pioneering studies usually adopt a two-stage pipeline, starting with a tokenizer converting POIs into discrete identifiers \addRed{that can be processed by LLMs}, followed by POI behavior prediction tasks to instruction-tune LLM for \addRed{next POI recommendation.}
Despite of remarkable progress, they still face two limitations: (1) \eat{the tokenizer struggles to encode heterogeneous signals from recommendation data, causing \textbf{information loss} during POI tokenization,}existing tokenizers struggle to encode heterogeneous signals in the recommendation data, suffering from information loss issue, and (2) \addRed{previous} instruction-tuning tasks only focus on users' POI visit behavior while ignore other behavior types, resulting in insufficient understanding of mobility. To address these limitations, we propose \textbf{\baby} (Knowledge Graph Tokenization for Behavior-aware generative next POI recommendation). Specifically, \baby organizes the recommendation data in a knowledge graph (KG) format, \addRed{of which the structure} can seamlessly preserve the heterogeneous information. Then, a KG-based tokenizer is developed to quantize each node into an individual structural ID. This process is supervised by the KG's structure, thus reducing the loss of heterogeneous information. 
Using generated IDs, \baby proposes multi-behavior learning that introduces multiple behavior-specific prediction tasks for LLM fine-tuning, e.g., POI, category, and region visit behaviors. Learning on these behavior tasks provides LLMs with comprehensive insights on the target POI visit behavior. Experiments on four real-world city datasets demonstrate the superior performance of \baby. The code and data are available in the Supplementary Material.
\end{abstract}

\input{introduction}

\input{relatedwork}
\input{problemstatement}

\input{Methodology}
\input{experiments}
\input{conclusion}

\bibliography{aaai2026}

\makeatletter
\def\isChecklistMainFile{}
\makeatother

\input{appendix}

\end{document}

%% file: introduction.tex
\section{Introduction}
With the widespread use of mobile devices, users are increasingly willing to share their geographic locations and experiences at points-of-interest (POIs) on the location-based social platforms. As a consequence, offering POI recommendation to users becomes a vital service in modern life, enabling users to explore and discover POIs that match their preferences. In the last decade, next POI recommendation--a subtask of POI recommendation--has been a popular research topic. This task aims at predicting a user's future movement in real-time based on his/her historical check-in records. 

Recently, in light of the impressive generative ability of Large Language Models (LLMs), generative paradigm has emerged as a new solution for next POI recommendation. Unlike conventional methods, generative methods redefine the next POI recommendation task as a sequence generation problem, which typically encompass two modules: \textbf{tokenization module} and \textbf{generation module}. In the tokenization module,  each POI is converted by a tokenizer into an individual discrete identifier that can be processed by LLMs.\eat{The identifier could be a random numeric ID (RID)~\cite{li2024large,wongso2024genup} or a sequence of discrete tokens, i.e., Semantic ID (SID)~\cite{chen2025enhancing}.}
Next, the generation module involves instruction-tuning, which prompts LLMs to autoregressively generate the identifiers of target POIs. Along this line, plenty of generative models have shown prominent performance~\cite{wang2023would,feng2024move,wang2025generative}.

\begin{figure}[tbp]
    \centering
    \includegraphics[width=0.48\textwidth]{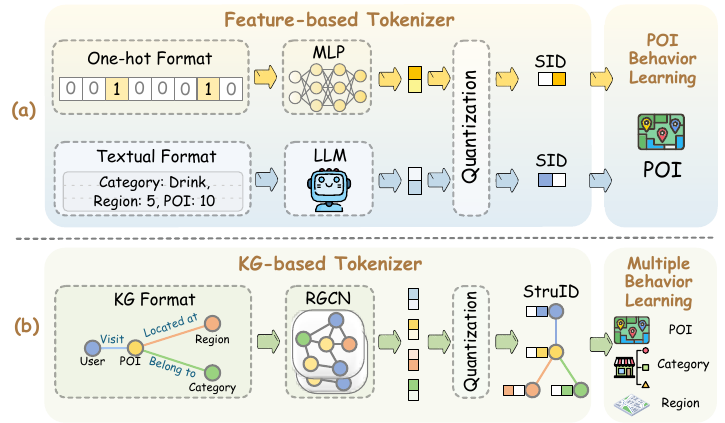}
    \caption{The comparison between previous methods and \baby. \addRed{(a) Previous methods that quantize POI features in the one-hot or textual format into SIDs. (b) \baby that organizes recommendation data in the KG format and quantizes nodes into StruIDs.}}
    \label{fig:introduction}
\end{figure}

Despite the effectiveness, pioneering methods face the following two limitations. (1) \textbf{Information loss during tokenization}. In POI recommendation data, there are heterogeneous types of information that facilitate recommendation, such as collaborative, geographical, and category signals~\cite{lai2024disentangled,lian2020geography,yu2020category}. Early ID-based tokenizers simply assign random numeric IDs (RIDs) to POIs\eat{~\cite{li2024large,wongso2024genup}}, inherently unable to incorporate heterogeneous information. Recent advanced feature-based tokenizers quantize one-hot or textual representations of POI features into semantic IDs (SIDs), as shown in Fig.~\ref{fig:introduction} (a), which contain rich information. However, they equally treat various heterogeneous signals and roughly compress them into a uniform representation for subsequent quantization, inevitably causing information loss.
(2) \textbf{Insufficient mobility understanding
during generation}. A user's visit to a POI indicates more than just an interaction with that specific location, it further reflects engagement with the category and region to which the POI belongs.
These auxiliary behavior types of interactions can offer richer insights into users' POI visit behavior.
However, the existing task prompts for instruction-tuning LLMs only focus on users' interactions with POIs, without considering other essential behavior types~\cite{wang2025generative,li2024large} that effect users' decisions on the next movements. This limits LLMs' ability to fully comprehend user mobility.

To address these limitations, we propose KGTB (Knowledge Graph Tokenization for Behavior-aware generative
next POI Recommendation). \addRed{For addressing the \underline{first limitation}, KGTB constructs a KG to organize the recommendation data, and develops a novel KG-based tokenizer to generate a new type of identifier named as structural ID (StruID), as showin in Fig.~\ref{fig:introduction} (b). Compared with existing one-hot and textual formats, the KG format seamlessly preserves heterogeneous data into its structure with different nodes and relations. Such structure information will be subsequently encoded into each node's StruID, achieved by two critical mechanisms of the KG-based tokenizer. One is the relational graph convolution network (RGCN)~\cite{schlichtkrull2018modeling} over the KG to create structure-aware representations for quantization. The other one is the KG reconstruction objective to supervise the representation quantization with link labels.}
For addressing the \underline{second limitation}, we propose multi-behavior learning to promote LLMs’ understanding of mobility. Specifically, we introduce three behavior-specific prediction tasks for LLM fine-tuning, with each behavior type (POI, category, or region visit behavior) corresponding to a prediction task. These tasks enable LLMs to comprehend  user mobility from diffrent views, resulting more accurate next POI recommendation.

In summary, this research has four contributions to the literature.
\begin{itemize}
    \item We propose a new method \baby for generative next POI recommendation, which activates the LLMs' understanding of user mobility via modeling heterogeneous information and learning multiple user behaviors.
    \item \addRed{We propose a novel KG-based tokenizer, which constructs a new type of identifier StruID incorporating the heterogeneous information preserved in the KG structure.} 
    To our best of knowledge, this is the first attempt to explore the potential of graph tokenizers for generative POI recommendation.
    \item We propose multi-behavior learning tasks for instruction-tuning LLMs, enhancing its understanding of user mobility from various views of behaviors.
    \item We conducted extensive experiments on four real-world LBSN datasets to evaluate the performance of \baby. Results show that \baby consistently outperforms existing conventional methods and generative methods
\end{itemize}

%% file: relatedwork.tex
\section{Related Work}
We divide related work into two categories: conventional methods and generative methods.
\subsection{Conventional Next POI Recommendation}
Conventional next POI recommendation methods follow the matching-paradigm. They encode user trajectories with sequence processing techniques and retrieve the most relevant POI from candidates for recommendation.
\eat{The key of next POI recommendation lies in effectively capturing the sequential patterns in user check-in history, hence the progress of conventional matching methods mirrors the development of sequence processing technology.}
Early methods employ Markov Chain (MC) to model naive transition patterns between consecutively visited POIs\eat{under geographical or temporal constraints}~\cite{cheng2013you,feng2015personalized}. 
With the development of recurrent and attention networks, more complex sequential patterns could be captured, using spatio-temporal information to control hidden state updates or attention scores~\cite{liu2016predicting,lian2020geography}.
\eat{Thereafter, the advent of recurrent neural networks and attention networks has made it feasible to model complex sequential patterns within long check-in history. To tailor these networks for next POI recommendation, researchers usually utilize the spatio-temporal information to control the update of hidden states~\cite{kong2018hst,liu2016predicting,zhao2019go}, or to assist in the calculation of attention scores~\cite{luo2021stan,zhang2022next,lian2020geography,wang2022spatial}.}\eat{For example, STGN~\cite{zhao2019go} modifies the gate mechanisms in LSTM by designing both temporal and spatial gates to control the impact of historical sequence on the current state. STAN~\cite{luo2021stan} modifies the attention mechanism in Transformer by leveraging time interval and geographical distance information.}
Subsequently, graph convolution networks become prevalent as graphs are an effective means to preserve global information~\cite{rao2024next,wang2023adaptive,lai2024disentangled}.
\eat{spatio-temporal information and incorporate global transition patterns from all users}
Propagation over constructed graphs enhances representations with collaborative signals from other related users or POIs, promoting recommendation performance.

Different from the conventional matching paradigm, in this paper we study a new generative paradigm for next POI recommendation, where the recommended POI is generated token by token.

\subsection{Generative Next POI Recommendation}
Recently, LLM-driven generative paradigm has shown promising performance in recommender systems, which autoregressively generates the identifiers of target POIs~\cite{zhu2024cost}.
\eat{Unlike the conventional matching paradigm, generative paradigm innovatively reformulates the recommendation task as a sequence generation problem, which autoregressively generates the identifiers of recommended targets}
This new paradigm has brought multifold merits, such as excellent scalability and lower computational redundancy~\cite{han2025mtgr}. 

For next POI recommendation, earlier studies rely on RIDs and sophisticated prompts to activate LLMs for next POI recommendation~\cite{li2024large,wongso2024genup,feng2024move,wang2023would}. They usually assign each POI a RID, then transform a user's historical movements into textual format with prompts, and finally take the text data as inputs to LLMs for recommendation results. Considering that RIDs cannot provide any semantic information about POIs, which largely limit the potential of LLMs, recent studies~\cite{wang2025generative,chen2025enhancing} propose to tokenize POIs into SIDs. Each SID is an ordered sequence of discrete tokens derived from a crafted POI feature vector or textual description. Similar SIDs usually denote correlated POIs, making it much easier for LLMs to understand user historical POI visits during generation.

However, existing methods suffer from the problems of information loss during tokenization and insufficient mobility understanding during generation.

\eat{Furthermore, when combined with prevalent LLMs, generative recommenders can be endowed with the abilities of reasoning~\cite{yang2025sparse} and semantic understanding~\cite{wang2024content}.}

%% file: problemstatement.tex
\section{Problem Statement}
\begin{figure*}[!h]
    \centering
    \includegraphics[width=\textwidth]{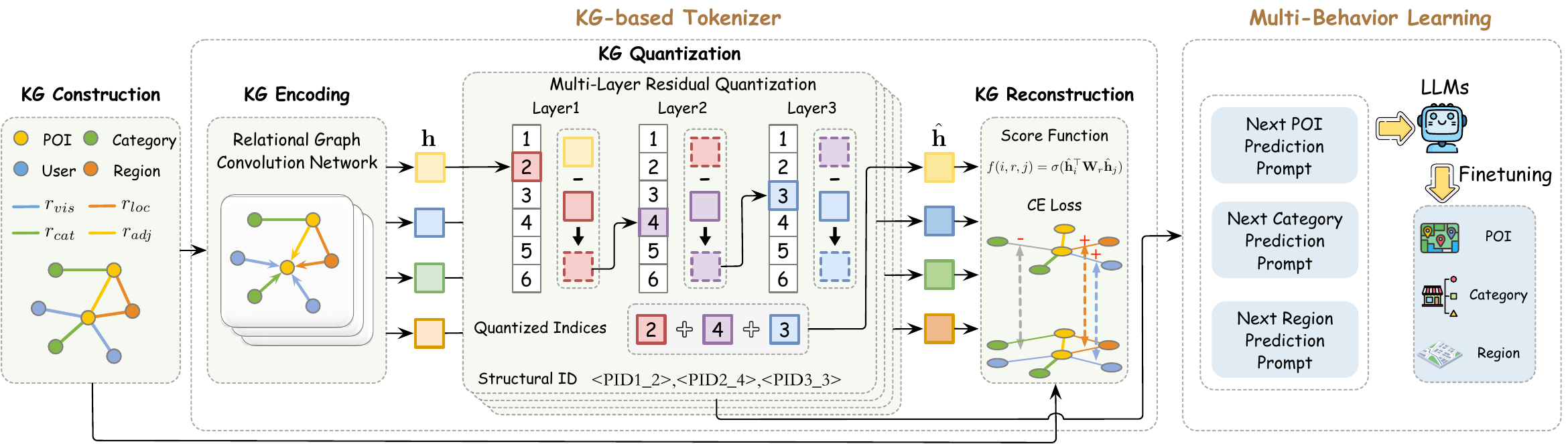}
    \caption{An overview of \baby method.}
    \label{fig:model}
\end{figure*}
We formulate the problem of next POI recommendation. Let $\mathcal{U}$ be a set of users, $\mathcal{V}$ be a set of POIs. Each POI $v_i\in\mathcal{V}$ is associated with a category $c_i\in\mathcal{C}$ and a region $o_i\in\mathcal{O}$. $\mathcal{C}$ and $\mathcal{O}$ are predefined category set and region set respectively. For each user $u\in\mathcal{U}$, his/her historical check-in sequence is denoted by $S^u=[s^u_1, s^u_2, ...,s^u_k]$, where each check-in tuple $s^u_i=(v_i,c_i,o_i,t_i)$ indicates that user $u$ visited POI $v_i$ characterized by category $c_i$ and region $r_i$ at timestamp $t_i$. Subsequently, we denote the historical category and region sequences of the user $u$ as $S^u_c=[c_1, c_2, ...,c_k]$ and $S^u_o=[o_1, o_2, ...,o_k]$, respectively. The next POI recommendation is defined as: Give a user $u$ and his/her check-in history $S^u$, the goal is to recommend a suitable POI $v_{k+1}$, one that $u$ is inclined to visit next.

%% file: Methodology.tex
\section{Methodology}

In this section, we present a detailed demonstration of our method \baby. As shown in Fig.~\ref{fig:model}, \baby first constructs a KG to retain heterogeneous information via its structure. Then \baby develops a KG-based tokenizer to generate StruIDs. Finally, \baby leverages StruIDs and defines multiple behavior-specific learning tasks to activate LLMs' ability for generative next POI recommendation.

\subsection{Knowledge Graph Construction}
\addRed{In this paper, we abandon the widely-used one-hot and textual formats. For the first time, we organize the recommendation data in a KG format for generative POI recommendation due to its innate ability to store heterogeneous data}.
Specifically, we construct KG $\mathcal{G}$, defined as $\left\{(h,r,t)|h,t\in{\mathcal{E}},r\in{\mathcal{R}}\right\}$. $\mathcal{E}$ denotes the entity set, which contains four sets of recommendation elements: user set $\mathcal{U}$, POI set $\mathcal{V}$, category set $\mathcal{C}$, and region set $\mathcal{O}$. Based on these entities, we build four types of relations $\mathcal{R}$. (1) \emph{Visit relation} $r_{vis}$ preserves collaborative signals represented by triple $(u_i,r_{vis},v_j)$, indicating user $u_i$ visited POI $v_j$. (2) \emph{Adjacent relation} $r_{vis}$ connects two geographically adjacent POIs $v_i$ and $v_j$ with distance lower than $d$ km, i.e., $(v_i,r_{adj},v_j)$, which delivers geographical signals. (3) \emph{Categorized relation} $r_{cat}$ links POI $v_i$ and its associated category $c_j$, i.e., $(v_i,r_{cat},c_j)$, retaining category signals. (4) \emph{Located relation} $r_{loc}$ connects POI $v_i$ and its located region $o_j$, represented by a triple $(v_i,r_{loc},o_j)$, preserving geographical signals. 

\eat{\begin{itemize}
	\item \emph{Visit relation} $r_{vis}$. If a user $u_i$ visited a POI $v_j$, then the nodes of $u$ and $v$ can be linked by a visit relation, represented by a triple $(u_i,r_{vis},v_j)$. This relation delivers collaborative signals. 
	\item \emph{Adjacent relation} $r_{adj}$. If a POI $v_i$ is geographically adjacent to another POI $v_j$ with distance smaller than $d$ km, then the nodes of $v_i$ and $v_j$ can be linked by an adjacent relation, represented by a triple $(v_i,r_{adj},v_j)$. This relation delivers geographical signals.
	\item \emph{Categorized relation} $r_{cat}$. If a POI $v_i$ belongs to a category $c_j$, then the nodes of $v_i$ and $c_j$ can be linked by a categorized relation, represented by a triple $(v_i,r_{cat},c_j)$. This relation delivers category signals. 
	\item \emph{Located relation} $r_{loc}$. If a POI $v_i$ is located at a region $o_j$, then the nodes of $v_i$ and $o_j$ can be linked by a located relation, represented by a triple $(v_i,r_{loc},o_j)$. This relation also delivers geographical signals. 
\end{itemize}}

The constructed $\mathcal{G}$ has preserved heterogeneous types of signals via its structure, i.e., relations, providing comprehensive information for the next tokenization procedure.

\subsection{Knowledge Graph-based Tokenizer}
As a crucial module in generative recommender systems, tokenizers quantize recommendation elements (e.g., users and POIs) into individual discrete identifiers \addRed{that can be processed by LLMs}. Prior studies implement tokenization by first encoding a POI's features into a semantic vector, and then quantizing it into the SID corresponding to that POI. However, equally dealing with different types of heterogeneous information and simply compressing them together into a single vector will cause inevitable information loss in SIDs.\eat{Additionally, they are unable to collaboratively tokenize multiple types of elements in the POI recommendation scenario, such as users, POIs, categories, and regions.}
To address this issue, we propose a KG-based tokenizer. It encodes the constructed KG to generate structure-aware representations, and quantizes them into StruIDs that incorporate abundant heterogeneous information stored in the KG's structure.


\subsubsection{Knowledge Graph Encoding}
Given $\mathcal{G}$, the KG-based tokenizer first encodes each node into a structure-aware representation, which will be used for subsequent quantization. We introduce RGCN~\cite{schlichtkrull2018modeling} to achieve the purpose. RGCN is an extension of GCNs that applies to relational data. The relation-aware information propagation of RGCN guarantees the efficient capture of structure information.


Each RGCN layer aggregates neighbors with relation-specific weight matrices to update node hidden representations. The aggregation operation is defined as follows:
\begin{equation}
    \mathbf{h}^{(l)}_i=\sigma(\mathbf{W}^{(l)}_0\mathbf{h}^{(l-1)}_i+\sum_{r\in\mathcal{R}}\sum_{j\in\mathcal{N}_r(i)}\frac{1}{|\mathcal{N}_r(i)|}\mathbf{W}^{(l)}_r\mathbf{h}^{(l-1)}_j),
\end{equation}
where $\mathbf{W}^{(l)}_0$ is the weight matrix for self-loop and $\mathbf{W}^{(l)}_r$ is the specific weight matrix for relation $r$. The set $\mathcal{N}_r(i)$ denotes the neighbors of node $i$ connected by relation $r$. $\mathbf{h}^{(0)}_i$ is node $i$'s randomly initialized embedding. By stacking $M$ RGCN layers, outputs of the last layer will be considered as structure-aware representations for later quantization. This procedure can be concisely formulated as:
\begin{equation}
    \mathbf{h}_{i}={\rm Encoder}(\mathcal{G},i),
\end{equation}
where $\mathbf{h}_{i}$ is the structure-aware representation for node $i\in\mathcal{E}$, integrating its local structure information.

\subsubsection{Knowledge Graph Quantization} 
After obtaining structure-aware representations, we then quantize them into StruIDs that can be processed by LLMs. A StruID is an ordered sequence of discrete tokens, and StruIDs with similar prefixes imply correlated entities in $\mathcal{G}$ with similar structures, e.g., connected to the same region or the same user. In practice, we adhere to the multi-layer residual quantization structure of RQVAE. \addRed{For each quantization layer, we allocate a codebook and use the residual output from the last layer as a query to retrieve the closed code from the current codebook. Output codes from all layers are combined to form a StruID.} 


\eat{To achieve this goal, we propose a knowledge graph quantization method. Specifically, we first calculate semantic vectors for all entities by applying relational graph convolution network~\cite{schlichtkrull2018modeling} (RGCN) over $\mathcal{G}$. The information propagation of RGCN guarantees the capture of high-order feature interactions among connected entities. To map semantic vectors into SIDs, we then modify the RQVAE structure with a proposed graph structure reconstruction objective, forcing SIDs to preserve the heterogeneous information stored in $\mathcal{G}$'s structure.}

\eat{After obtaining semantic vectors, we now quantize them into discrete semantic IDs following the pipeline of RQVAE,}

The first step is to establish codebooks. Considering the different cardinal numbers of the four entity types, we maintain an individual hierarchical codebook set for each type. Formally, for entity type $e\in\left\{ user, POI, category, region\right\}$, we define the associated codebook set as:
\begin{equation}
    \mathcal{B}^e=\left\{\mathbf{B}^e_1,\mathbf{B}^e_2,...,\mathbf{B}^e_{L}\right\},
\end{equation}
where $\mathbf{B}^e_l\in\mathbb{R}^{K\times d}$ denotes a codebook of size $K$ at the $l$-th quantization layer. $L$ represents the number of quantization layers. 

With the codebook set $\mathcal{B}^e$, the StruID for node $i$ of type $e$ can be generated recursively in a coarse-to-fine manner. We first set initial residual $\mathbf{z}_1$ as the structure-aware vector $\mathbf{h}_{i}$. Then we take $\mathbf{z}_1$ as input to the first quantization layer, and perform a nearest neighbor search on codebook $\mathbf{B}^e_1\in\mathcal{B}^e$:
\begin{equation}
    \label{eq:index}
    n^e_1={\rm arg}\,{\rm min}_{k}\|\mathbf{z}_1-\mathbf{b}^e_{1,k}\|,
\end{equation}
where $\mathbf{b}^e_{1,k}$ is the code vector from $\mathbf{B}^e_1$. The resulted $n^e_1$ denotes the index of the closet code vector in $\mathbf{B}^e_1$, which will be treated as the 1-th token in node $i$'s StruID. For the rest $l$-th token, we conduct the operation of Eq.~\ref{eq:index} at the $l$-th quantization layer with residual $\mathbf{z}_{l}$ as input, which is defined as:
\begin{equation}
    \label{eq:residul}
    \mathbf{z}_{l}=\mathbf{z}_{l-1}-\mathbf{b}^e_{l-1,n^e_{l-1}},\,l\in[2,L],
\end{equation}
Here, $\mathbf{z}_{l-1}$ and $n^e_{l-1}$ represent the input residual and the output code index at the ($l-1$)-th layer, respectively. $\mathbf{z}_{l}$ encodes the leftover information that has not been recognized by the ($l-1$)-th layer.

Through $L$ layers of residual quantization, we obtain a code of quantized indices $\mathbf{n}_i=[n^e_1,n^e_2,...,n^e_L]$ that formulates the StruID for node $i$. 



\subsubsection{Knowledge Graph Reconstruction}
To ensure StruIDs learn heterogeneous information stored in $\mathcal{G}$'s structure, we propose to reconstruct $\mathcal{G}$ using the quantized indices $\mathbf{n}_i$. This distinguishes our method from RQVAE. The original objective of RQVAE is to reconstruct the input feature vector. However, in our scenario, it is hard to fully express a KG via vectors due to its heterogeneity, making RQVAE unsuitable.

Specifically, given node $i$ and its quantized indices $\mathbf{n}_i$, we first decode it into  quantized representation $\hat{\mathbf{h}}_i=\sum^{L}_{l=1}\mathbf{b}^e_{l,n^e_{l}}$. Then we define a score function to measure the possibility of node $i$ being connected to another node $j$ with relation $r$:
\begin{equation}
	f(i,r,j)=\sigma(\hat{\mathbf{h}}^{\top}_i\mathbf{W}_r\hat{\mathbf{h}}_j),
\end{equation}
where $\hat{\mathbf{h}}_j$ is the quantized vector for node $j$ and $\mathbf{W}_r$ denotes the weight matrix for relation $r$. Finally, we introduce a new KG reconstruction objective $\mathcal{L}_{KG}$, which scores higher to positive triples from $\mathcal{G}$ than negative ones. In practice, $\mathcal{L}_{KG}$ is formulated as a cross-entropy (CE) loss:
\begin{equation}
	\mathcal{L}_{KG}=y{\rm log}(f(i,r,j))+(1-y){\rm log}(1-f(i,r,j)),
\end{equation}
where $y$ is the link label of whether triple $(i,r,j)$ is valid, which is 1 if the triple is observable in $\mathcal{G}$, 0 otherwise.

\subsubsection{Training} The training of our KG-based tokenizer consists of two objectives. One is our proposed $\mathcal{L}_{KG}$, the other one comes from RQVAE, which encourages input residuals to cluster around their closet code vectors in codebooks:
\begin{equation}
	\mathcal{L}_{RQ}=\sum^L_{l=1}\|{\rm sg}[\mathbf{z}_l]-\mathbf{b}^e_{l,n^e_l}\|^2+\beta\|{\rm sg}[\mathbf{b}^e_{l,n^e_l}]-\mathbf{z}_l\|^2,
\end{equation}
where ${\rm sg}$ is the stop-gradient operation, $\beta$ is a balance factor. Finally, the overall objective is defined as $\mathcal{L}_{KG}+\mathcal{L}_{RQ}$.

\subsection{Multi-Behavior Learning}
The KG-based tokenizer has converted all recommendation elements into StruIDs. Leveraging these StruIDs, we construct multiple behavior-specific tasks to instruction-tune LLMs for generative next POI recommendation. In the literature, most tasks used in existing studies only involve users' interactions with POIs~\cite{wang2025generative,chen2025enhancing}, while overlook other critical behavior types, i.e., interactions with categories and regions. In reality, a user's visit decision on a POI usually begins by considering about the category based on his/her personal intent, followed by choosing the region according to geographical conditions.
Hence, learning from these auxiliary category and region behaviors enables LLMs a comprehensive understanding of user mobility. To this end, we propose multi-behavior learning with behavior-specific tasks, including one main task for predicting next POI and two auxiliary tasks for predicting next category and region.

\subsubsection{Next POI Prediction Prompting} Given a user $u\in\mathcal{U}$ and his/her historical check-in sequence $S^u$, we prompt LLMs to generate the next POI $v_{k+1}$. Specifically, we construct a next POI prediction prompt to convert the task into a natural language format, which is composed of an input block and a target block. The input block mainly contains $u$'s POI preference and check-in history. The POI preference\footnote{In practice, we represent the user's POI preference with his/her top-5 most frequently visited POIs.} reflects the overall tendency of the user to visit POIs, while the check-in history delivers essential sequential contexts. Both of them contribute to an accurate next POI generation. In the target block, ground-truth POI $v_{k+1}$ is included as the label to fine-tune LLMs. An example prompt is shown below:
\begin{tcolorbox}[boxrule=0.4pt,colback=white,width=0.45\textwidth,height=0.2\textwidth]
{\small \textbf{Input Block}: Please conduct a next POI recommendation. There is user \colorbox{yellow_our}{UID} and his preferable POIs: \colorbox{blue_our}{CID$_1$}\colorbox{purple_our}{OID$_1$}\colorbox{pink_our}{PID$_1$},.... Here is his current trajectory: visiting \colorbox{blue_our}{CID$_1$}\colorbox{purple_our}{OID$_1$}\colorbox{pink_our}{PID$_1$} at time $t_1$, ..., visiting \colorbox{blue_our}{CID$_2$}\colorbox{purple_our}{OID$_2$}\colorbox{pink_our}{PID$_{2}$} at time $t_2$. Which POI will the user \colorbox{yellow_our}{UID} visit at time $t_{3}$?
\\
\textbf{Output Block}: POI \colorbox{pink_our}{PID$_{3}$}}
\end{tcolorbox}
Note that \colorbox{yellow_our}{UID}, \colorbox{blue_our}{CID}, \colorbox{purple_our}{OID}, and \colorbox{pink_our}{PID} are user, category, region, and POI StruIDs, respectively.

\subsubsection{Next Category and Region Prediction Prompting} Given a user $u\in\mathcal{U}$ and his/her historical category sequence $S^u_c$ and region sequence $S^u_o$, LLMs are expected to generate the next POI $v_{k+1}$'s associated category $c_{k+1}$ and region $r_{k+1}$. For each prediction task, we build the prompt for natural language conversion in a similar way to the POI prediction prompt.

Based on the designed prompts, we convert the check-in data into textual data for LLM fine-tuning. In order to evaluate the fine-tuned LLM, we use the next POI prediction prompt for next POI recommendation.

\eat{These two auxiliary tasks enable LLMs to comprehend the user's category and region behaviors, providing complementary insights on the target POI behavior. For each of them, the prediction prompt is constructed in a similar way to the POI prediction prompt. The input template includes the user's preferences and history, while the target template is the ground-truth.}

\eat{
which pushes node $i$'s quantized vector approximate to its neighbors.  
scores positive triplets higher than the negative ones: 
Assume two nodes $i$ and $j$ connected with the relation $r$ in the KG, $\mathcal{L}_{KG}$ is calculated as 
we first decode their codes into quantized vectors $\hat{\mathbf{h}}_i$ and $\hat{\mathbf{h}}_j$:
In our case, however, the original representation reconstruction objective in RQVAE is not suitable anymore as we intend to learn structure information stored in $\mathcal{G}$. 
Meanwhile, the node $i$'s semantic vector $\mathbf{h}_i$ is also quantized into   
With the codebook $\mathbf{B}^e_i$ at the $i$-th layer, 
A straightforward way would be directly employ the RQVAE over these vectors. In our case, however, the original representation reconstruction objective in RQVAE is not suitable anymore as we intend to learn structure information stored in $\mathcal{G}$.}

%% file: experiments.tex
\section{Experiments}

\eat{In this section, we conduct extensive experiments to evaluate the performance of our \baby.}
\subsection{Experimental Setup}
\subsubsection{Datasets and Metrics}
We evaluate the performance of our \baby on two widely-used global-scale datasets: Foursquare~\cite{yang2016participatory} and Gowalla~\cite{cho2011friendship}.
\eat{The Foursquare dataset~\cite{yang2016participatory} spans from April 2012 to September 2013, while the Gowalla dataset~\cite{cho2011friendship} includes check-in records from February 2009 to October 2010.} 
From each dataset, we choose two representative cities for experiments, ultimately forming four city-scale datasets: \textbf{Foursquare-Boston (FB)}, \textbf{Foursquare-Paris (FP)}, \textbf{Gowalla-Austin (GA)}, and \textbf{Gowalla-San Francisco (GS)}. The statistics for these city datasets are presented in Tab.~\ref{tab:datastatistics}. 
For partition, we split each user's records chronologically into 70\% for training, 10\% for validation, and 20\%
for testing.
\begin{table}[h]
	\footnotesize
	
	\centering
	\setlength\tabcolsep{2.0pt}
	\begin{tabular}{ccccc}
		\toprule
		Dataset &\#User &\#POI &\#Record &\#Category  \\
		\midrule
		Foursquare-Boston (FB)&4,595  &14,445  &104,181&398\\
		Foursquare-Paris (FP)&6,904  &19,838  &111,277&409\\
		\midrule
		Gowalla-Austin (GA)&7,841  &18,252  &317,199&369\\
		Gowalla-San Francisco (GS)&6,105  &10,631  &144,505&285\\
		\bottomrule
	\end{tabular}
    \caption{Statistics of four city datasets.}
    \label{tab:datastatistics}
\end{table}

To evaluate recommendation performance, we adopt two widely utilized metrics: Hit Rate at a cutoff top K (HR@K) and Normalized discounted cumulative gain at a cutoff top K (N@K). K is selected from $\left\{1,5,10\right\}$.

\subsubsection{Baselines}
We compare \baby with nine baseline methods for next POI recommendation, which can be categorized into four groups. (1) \textbf{RNN-based method}: STLSTM~\cite{kong2018hst} redefines LSTM's gate mechanism with temporal and spatial information. (2) \textbf{Attention-based method}: SASRec~\cite{zhang2018next} utilizes the self-attention network. (3) \textbf{Graph-based methods}: GETNext~\cite{yang2022getnext} captures global collaborative signals from a global graph that aggregates all users' trajectories. HMTGRN~\cite{lim2022hierarchical} constructs spatio-temporal graphs to learn POI-POI relationships. Graph-FlashBack~\cite{rao2022graph} and ARGAN~\cite{wang2023adaptive} enhance representations via adaptively refined POI graphs. (4) \textbf{Generative methods}: P5~\cite{geng2022recommendation} unifies recommendation tasks in a shared sequence-to-sequence format. LLM4POI~\cite{li2024large} fine-tunes LLMs for next POI recommendation with trajectory prompting. GNPR-SID~\cite{wang2025generative} tokenizes POIs into spatio-temporal-aware SIDs for a more efficient LLM fine-tuning. 

\subsubsection{Implementation Details} For KG construction, the \addRed{distance threshold $d$} is set to 0.2. For the KG-based tokenizer, the layer number for RGCN and quantization is set to 3. For the $l$-th quantization layer, $l\in[1,2,3]$, the codebook size is set to $8^{4(l-1)}$ for POI and user entities, $8^{2(l-1)}$ for region and category entities, with each code of 64 dimensions. The balance factor $\beta$ in $\mathcal{L}_{RQ}$ is set to 0.25. For multi-behavior learning, we choose a small decoder-only model GPT-2 as our backbone LLM to meet the low latency requirement in recommender systems. The fine-tune process is conducted on an NVIDIA L40 GPU for 8 epochs, using LoRA~\cite{hu2022lora} with a rank of 16, a learning rate of 0.0002, a batch size of 4. We repeat the experiment 3 times.

\begin{table*}[t]
	\small
	\setlength{\tabcolsep}{2.5pt}
	\centering
	\begin{tabular}{c|c|ccccccccc|cr}
		\toprule
		\textbf{Dataset} & \textbf{Metric} & \textbf{STLSTM} & \textbf{SASRec} & \textbf{GETNext} & \textbf{\makecell{Graph\\FlashBack}} & \textbf{HMTGRN} & \textbf{ARGAN} & \textbf{P5} & \textbf{LLM4POI} & \textbf{GNPR-SID} & \textbf{\baby} & \textbf{Imp.} \\
		\midrule
		\multirow{5}[0]{*}{\makecell[c]{FB}}
              &HR@1&0.1119&0.1212&0.1061&0.1384&0.1199&0.1232&0.1392&0.0566&\underline{0.1577}&\textbf{0.1675}&6.2\% \\
		     &HR@5&0.2118&0.2264&0.2139&\underline{0.2849}&0.2214&0.2594&0.2833&-&0.2827&\textbf{0.3190}& 12.0\% \\
		   &N@5&0.1646&0.1774& 0.1629&0.2157&0.1730&0.1951&0.2154&-&\underline{0.2231}&\textbf{0.2468}& 10.6\% \\
		   &HR@10&0.2531&0.2643&0.2622&\underline{0.3494}&0.2707&0.3119&0.3452&-&0.3231&\textbf{0.3818}& 9.3\% \\
              &N@10&0.1779&0.1898&0.1786&\underline{0.2366}&0.1890&0.2121&0.2355&-&0.2364&\textbf{0.2673}& 13.0\% \\
		\midrule
		\multirow{5}[0]{*}{\makecell[c]{FP}}
              &HR@1&0.0954&0.1023&0.1028&0.1314&0.1089&0.1236&0.1447&0.0329&\underline{0.1649}&\textbf{0.1716}&4.1\% \\
		     &HR@5&0.1857&0.1969&0.2035&0.2689&0.2061&0.2561&0.2823&-     &\underline{0.2858}&\textbf{0.3193}&11.7\% \\
		   &N@5&0.1433&0.1526& 0.1555&0.2038&0.1602&0.1933&0.2175&-     &\underline{0.2282}&\textbf{0.2485}& 8.9\% \\
		   &HR@10&0.2239&0.2330&0.2447&0.3284&0.2469&0.3089&\underline{0.3440}&-    &0.3271&\textbf{0.3789}& 10.1\% \\
              &N@10&0.1557&0.1642&0.1689&0.2231&0.1734&0.2105&0.2375&-     &\underline{0.2417}&\textbf{0.2679}& 10.8\% \\
		\midrule
		\multirow{5}[0]{*}{\makecell[c]{GA}}
              &HR@1 &0.0707&0.0756&0.0726&0.0829&0.0728&0.0614&0.0918&0.0209&\underline{0.1005}&\textbf{0.1085}&8.0\% \\
		     &HR@5 &0.1512&0.1571&0.1632&0.1908&0.1552&0.1560&\underline{0.1982}&-     &0.1890&\textbf{0.2250}& 13.5\% \\
		   &N@5  &0.1124&0.1181&0.1198&0.1392&0.1155&0.1104&\underline{0.1476}&-     &0.1466&\textbf{0.1693}& 14.7\% \\
		   &HR@10&0.1961&0.2012&0.2076&0.2429&0.2002&0.2072&\underline{0.2506}&-     &0.2244&\textbf{0.2744}& 9.5\% \\
              &N@10 &0.1269&0.1323&0.1341&0.1560&0.1300&0.1270&0.1476&-     &\underline{0.1582}&\textbf{0.1854}& 17.2\% \\
		\midrule
		\multirow{5}[0]{*}{\makecell[c]{GS}}
              &HR@1 &0.0740&0.0762&0.0731&0.0863&0.0768&0.0654&0.0924&0.0243&\underline{0.0911}&\textbf{0.1065}&16.9\% \\
		     &HR@5 &0.1604&0.1639&0.1635&0.1943&0.1648&0.1663&\underline{0.1971}&-     &0.1736&\textbf{0.2258}& 14.6\% \\
		   &N@5  &0.1191&0.1221&0.1202&0.1434&0.1227&0.1182&\underline{0.1480}&-     &0.1350&\textbf{0.1696}& 14.6\% \\
		   &HR@10&0.2012&0.2018&0.2101&\underline{0.2413}&0.2127&0.2137&0.2401&-     &0.1999&\textbf{0.2706}& 12.1\% \\
              &N@10 &0.1322&0.1343&0.1353&0.1587&0.1382&0.1335&\underline{0.1618}&-     &0.1435&\textbf{0.1842}& 13.8\% \\
		\bottomrule
	\end{tabular}%
	
    \caption{Performance comparison with baselines on four city datasets. For each row, the best results are highlighted in bold, and the second best scores are underlined. Imp. denotes the relative improvement ratio of \baby compared to the best baseline.}
    \label{tab:allperformance}
\end{table*}%

\subsection{Main Results}
We report the comparison results in Tab.~\ref{tab:allperformance} and derive the following three observations. 

Firstly, \baby consistently outperforms the best baselines across all evaluation metrics, with an average improvement of 10.22\%, 9.12\%, 12.58\%, 14.4\% on FB, FP, GA, and GS, respectively. These results demonstrate the superiority of our proposed \baby and its generalization across different domains and cities. 

Secondly, \baby significantly surpasses generative baseline methods. Compared with P5 and GNPR-SID, which adopt POI SIDs and produce the highest scores among baselines, \baby achieves a 13.0\% relative gain on HR@5, showing an outstanding generation capability for next POI recommendation. Unexpectedly, LLM4POI suffers from poor performance, which could be attributed to two factors. First, the RIDs employed by LLM4POI restrict LLMs' comprehension of semantic relationships among POIs. Second, LLM4POI relies on dense trajectory data as LLM input to infer user movements, which is not applicable to our sparser datasets~\footnote{LLM4POI removes short trajectories and filters out inactivate users and POIs, making datasets more denser. On the contrary, we avoid using any filter strategies to mimic real-world recommendation scenarios.}.

Lastly, in some cases, generative baseline methods are beaten by conventional methods (e.g., especially graph-based methods). In contrast, \baby is able to obtain superior performance than all conventional methods, demonstrating the effectiveness of our KG-based tokenizer and multi-behavior learning strategy.

We also conduct experiments with Llama3-8B as \baby's backbone. The experiments yield findings that are consistent with the aforementioned conclusions. The details are shown in the Appendix of Supplementary Material.


\eat{\begin{table}[h]
	\setlength{\tabcolsep}{0.5em}
	\centering
	
	\begin{tabular}{lcccc}
		\toprule
		\multirow{1}[0]{*}{Model} & \multicolumn{1}{c}{FB} & \multicolumn{1}{c}{FP} & \multicolumn{1}{c}{GA} & \multicolumn{1}{c}{GS}\\
		\midrule
		\textbf{Full}  & \textbf{0.1675} & \textbf{0.1716} & \textbf{0.1085} & \textbf{0.1065} \\
		\midrule
		\textbf{w/o SID} & 0.1631 $\downarrow$&   &   &  \\
		\textbf{w/o Reg} & 0.1660 $\downarrow$& 0.1683 $\downarrow$& 0.0919 $\downarrow$& 0.0913$\downarrow$\\
		\textbf{w/o Cat} & 0.1606 $\downarrow$& 0.1693 $\downarrow$& 0.0935 $\downarrow$& 0.0945$\downarrow$\\
		\textbf{w/o RegCat} & 0.1575 $\downarrow$& 0.1599 $\downarrow$& 0.0902 $\downarrow$& 0.0883$\downarrow$\\
            \textbf{w/o Pref} & 0.1575 $\downarrow$& 0.1599 $\downarrow$& 0.0902 $\downarrow$& 0.0883$\downarrow$\\
            \textbf{w/o Seq} & 0.1575 $\downarrow$& 0.1599 $\downarrow$& 0.0902 $\downarrow$& 0.0883$\downarrow$\\
		\bottomrule
	\end{tabular}%
        \caption{Ablation study on key components of \baby on HR@1. "Full" denotes the full model that produces the best results.}
	\label{tab:ablation}%
\end{table}}

\subsection{In-depth Analysis}
\subsubsection{Ablation Study}To analysis the effectiveness of each component in \baby, we conduct an ablation study on four city datasets with six variants: (1) \emph{w/o StruID} replaces StruIDs with RIDs. (2) \emph{w/o Reg} removes the region behavior prediction task. (3) \emph{w/o Cat} removes the category behavior prediction task. (4) \emph{w/o RegCat} removes both the region and category behavior prediction tasks. (5) \emph{w/o Pref} removes the preference from the prompt. (6) \emph{w/o Seq} removes the check-in history from the prompt. We report N@5 scores in Tab.~\ref{tab:ablation}.

\begin{table}[h!]
	\setlength{\tabcolsep}{0.5em}
	\centering
	
	\begin{tabular}{lcccc}
		\toprule
		Model & \multicolumn{1}{c}{FB} & \multicolumn{1}{c}{FP} & \multicolumn{1}{c}{GA} & \multicolumn{1}{c}{GS}\\
		\midrule
		\textbf{\baby}  & \textbf{0.2468} & \textbf{0.2485} & \textbf{0.1693} & \textbf{0.1696} \\
		\midrule
		\emph{w/o StruID} & 0.2396 $\downarrow$& 0.2441$\downarrow$  &0.1509  $\downarrow$ & 0.1496 $\downarrow$\\
		\emph{w/o Reg} & 0.2456 $\downarrow$& 0.2456 $\downarrow$& 0.1539 $\downarrow$& 0.1546$\downarrow$\\
		\emph{w/o Cat} & 0.2440 $\downarrow$& 0.2466 $\downarrow$& 0.1554 $\downarrow$& 0.1575$\downarrow$\\
		\emph{w/o RegCat} & 0.2411 $\downarrow$& 0.2421 $\downarrow$& 0.1529 $\downarrow$& 0.1523$\downarrow$\\
            \emph{w/o Pref} & 0.2410 $\downarrow$& 0.2368 $\downarrow$& 0.1494 $\downarrow$& 0.1425$\downarrow$\\
            \emph{w/o Seq} & 0.2088 $\downarrow$& 0.2050 $\downarrow$& 0.1328 $\downarrow$& 0.1350$\downarrow$\\
		\bottomrule
	\end{tabular}%
        \caption{Ablation study on key components of \baby.}
	\label{tab:ablation}%
\end{table}%

Based on the statistics from Tab.~\ref{tab:ablation}, we have the following observations. (1) Removing any component from \baby will cause a performance decrease, verifying the positive impact of each component. (2) \emph{w/o StruID} performs significantly worse than \baby. This demonstrates that our StruIDs provide the backbone model with valuable information captured from the constructed KG. (3) The removal of auxiliary behavior tasks hurts \baby's performance, showing that learning users' multiple behaviors is beneficial. (4) \emph{w/o Pref} and \emph{w/o Seq} suffer from an obvious drop on all datasets. This indicates that both of the user preference and the check-in history contribute to the accurate generation. 

\begin{figure*}[t]
	\centering
	\subfloat[StruIDs w.r.t region]{
		\includegraphics[width=0.24\textwidth,height=0.16\textwidth]{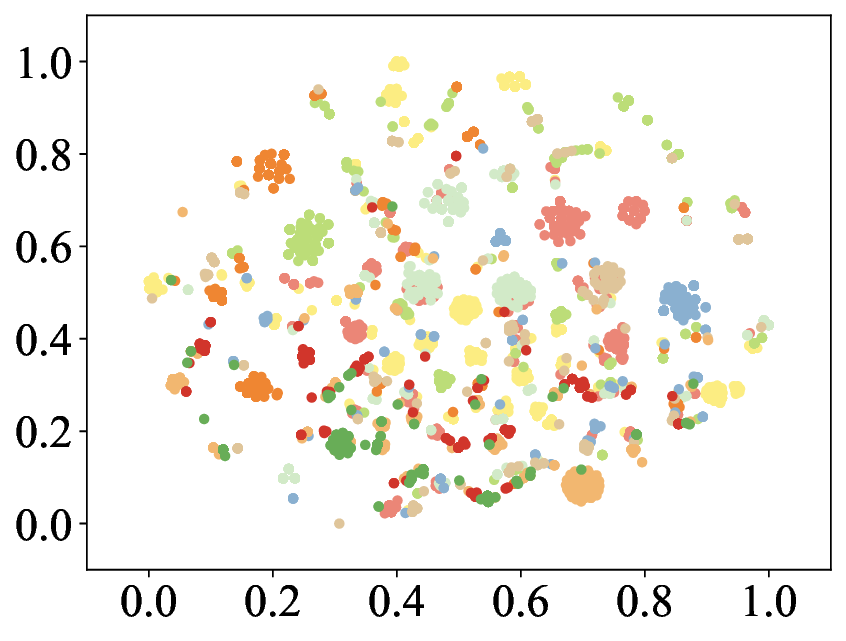}\label{fig:StruIDRegion}}
	\subfloat[SIDs w.r.t region]{
		\includegraphics[width=0.24\textwidth,height=0.16\textwidth]{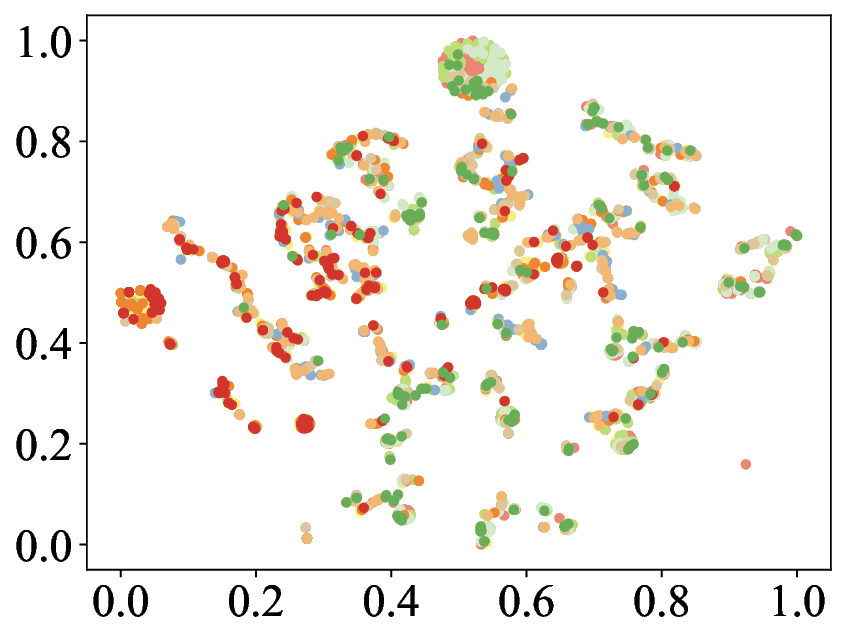}\label{fig:SIDRegion}} 
	\subfloat[StruIDs w.r.t category]{
		\includegraphics[width=0.24\textwidth,height=0.16\textwidth]{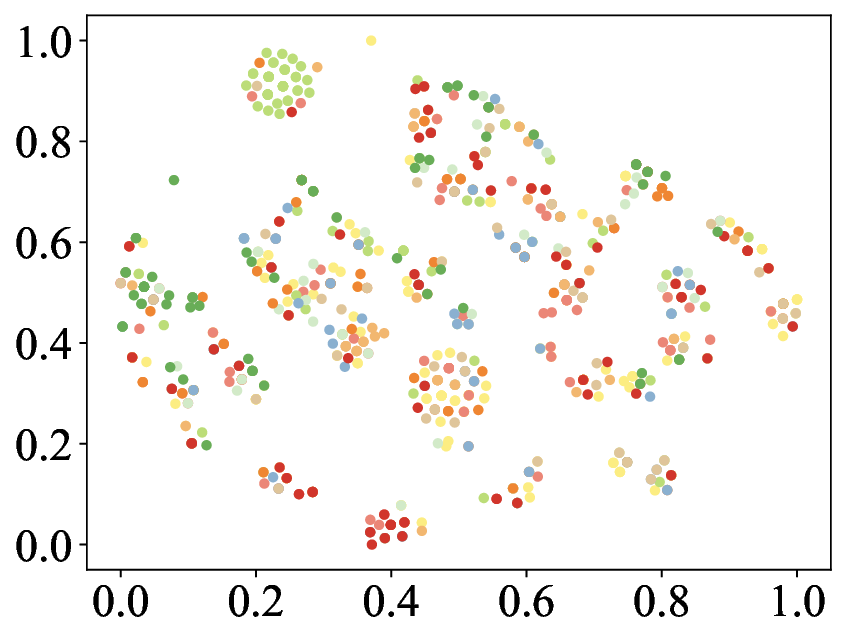}\label{fig:StruIDCategory}}
		\subfloat[SIDs w.r.t category]{
		\includegraphics[width=0.24\textwidth,height=0.16\textwidth]{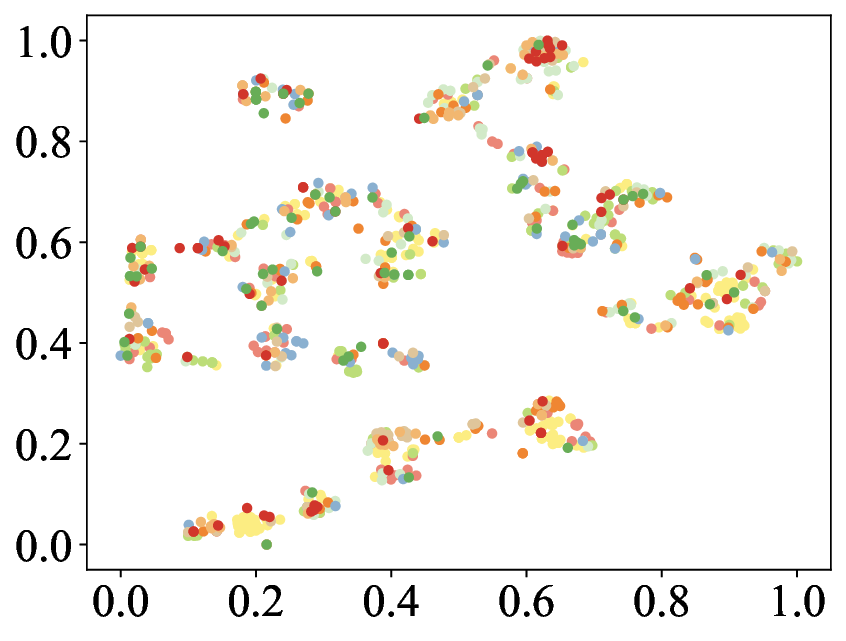}\label{fig:SIDCategory}}
	\caption{Visualization of StruIDs from \baby and SIDs from GNPR-SID on FB dataset. Circles of the same color belong to the same region or category. }\label{fig:emb}
\end{figure*}

\subsubsection{Performance on Cold-Start Scenario}
\begin{figure}[tbp]
    \centering
    \includegraphics[width=0.38\textwidth]{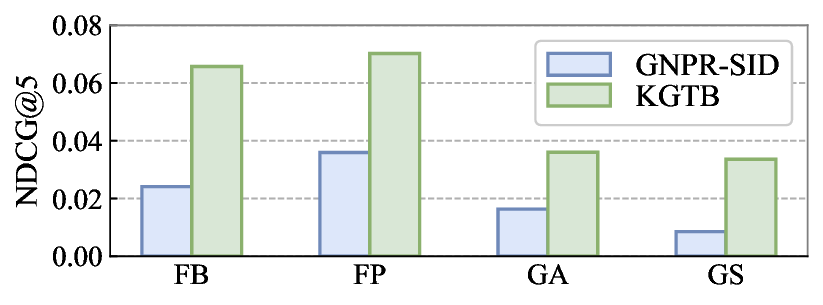}
    \caption{Performance comparison on cold-start POIs.}
    \label{fig:cold_start}
\end{figure}

POI StruIDs preserve rich heterogeneous information, which help alleviate POI cold-start problem. To verify this, we compare the performance of \baby and GNPR-SID on cold-start POIs with fewer than 5 visitors. We choose GNPR-SID for comparison in this and the rest of experiments because it is the most competitive baseline and the most relevant method to \baby. Results in Fig.~\ref{fig:cold_start} suggest that \baby achieves substantial performance gains over GNPR-SID across all datasets, highlighting its effectiveness in handling cold-start POIs.

\subsubsection{Performance on Out-of-Domain Data}
Generative recommendation methods benefit from the powerful generalization ability of the backbone LLMs. In this experiment, we analyze the performance of \baby on out-of-domain datasets and show the performance comparison between \baby and GNPR-SID in Tab.~\ref{tab:cross_city}. 

Results from Tab.~\ref{tab:cross_city} show that our approach \baby experiences a smaller performance decline on all out-of-domain datasets than GNPR-SID, suggesting a stronger generalization ability of \baby. We attribute this superiority to our proposed StruIDs, which learn rich structure information from the constructed KG. Training with StruIDs on source data endows \baby with the ability to recommend POIs in awareness of the KG structure, e.g., recommending POIs connected by the adjacent relation. Such ability can be easily generalized to those KGs in other fields, allowing \baby to produce better performance on out-of-domain data.

\begin{table}[h]
	\setlength{\tabcolsep}{0.4em}
	\centering
	
	\begin{tabular}{llcccc}
		\toprule
		\multirow{1}[0]{*}{Model} & \multirow{1}[0]{*}{Dataset}& \multicolumn{1}{c}{FB} & \multicolumn{1}{c}{FP} & \multicolumn{1}{c}{GA} & \multicolumn{1}{c}{GS}\\
		\midrule
        \multirow{4}[0]{*}{GNPR-SID} & FB &0.2231&0.0593&0.0480 &0.0559\\
             & FP &0.0870&0.2282&0.0259  &0.0502\\
          & GA &0.0576&0.0648&0.1466&0.0253 \\
             & GS &0.0458&0.0659&0.0250  &0.1350\\
             \midrule
		\multirow{4}[0]{*}{\baby} & FB &0.2468&0.1983&0.1167&0.1044\\
                                      & FP &0.1761&0.2485&0.1018&0.1025\\
                                      & GA &0.1685&0.1604&0.1693&0.1083\\
                                      & GS &0.1656&0.1465&0.1128&0.1696\\

		\bottomrule
	\end{tabular}%
        \caption{N@5 performance comparison on out-of-domain datasets. Datasets in the second column indicate the training domain, while those in the first row denote the test domain.}
	\label{tab:cross_city}%
\end{table}

\subsubsection{Analysis of StruID}
StruID encodes the structure of KG $\mathcal{G}$, so as to seamlessly preserve the heterogeneous information from the POI recommendation data, e.g., category and region. To verify whether or not StruID achieves this goal, we compare it with SID from GNPR-SID through visualization on FB dataset.

First, we choose top-10 regions that contain the most POIs, and visualize the StruIDs and SIDs of POIs belonging to these regions. Fig.~\ref{fig:emb} (a) and Fig.~\ref{fig:emb} (b) present the visualization results for StruID and SID respectively. Apparently, StruIDs are prone to group together by region, while this phenomenon is not obvious for SIDs. Then, we select top-10 frequent categories from previous regions and do the visualization again. Fig.~\ref{fig:emb} (c) and Fig.~\ref{fig:emb} (d) show the results. Although StruIDs do not form tight clusters by category as they do by region, we still find some noticeable clusters, e.g., in the upper left of Fig.~\ref{fig:emb} (c). In contrast, SIDs in Fig.~\ref{fig:emb} (d) are more scattered. To conclude, our proposed StruID is able to integrate expected heterogeneous information.

\subsubsection{Analysis of Model Efficiency} We analyze the computational cost of \baby by comparing it to GNPR-SID on FB dataset and show the results in Tab.~\ref{tab:cost}. \baby with GPT2 as the backbone has a much lower demand for resources than GNPR-SID based on more effective Llama3-8B, but excels in recommendation performance. This further demonstrates the effectiveness of our proposed method.

\begin{table}[h]
	\setlength{\tabcolsep}{0.5em}
	\centering
	\begin{tabular}{lccc}
		\toprule
		Model & \multicolumn{1}{c}{Training Time} & \multicolumn{1}{c}{Test Time} & \multicolumn{1}{c}{Parameter Size}\\
		\midrule
        GNPR-SID &39.3h&6h&1,072M\\
		\baby  &5.5h &0.4h&80.2M\\
		
        \midrule
	\end{tabular}%
        \caption{Comparison of computational cost on FB dataset.}
	\label{tab:cost}%
\end{table}%

%% file: conclusion.tex
\section{Conclusion}
In this paper, we propose a new method, namely \baby, for generative next POI recommendation. \baby organizes the recommendation data in a KG format to preserve heterogeneous information. Based on the constructed KG, \baby introduces a novel KG-based tokenizer to quantize all entities into structural IDs under the assistance of relational graph convolution network and KG reconstruction objective. These IDs seamlessly integrate heterogeneous information stored in the KG's structure, facilitating the recommendation performance and generalization capability of the backbone LLM. Furthermore, to provide a more comprehensive understanding of mobility, \baby involves multi-behavior learning, which fine-tunes the LLM with multiple behavior-specific prediction tasks. Experiments on four real-world datasets and across various scenarios indicate the superiority of \baby in terms of recommendation accuracy, cold-start problem handling, out-of-domain generalization, and computational efficiency.

%% file: appendix.tex
\tcbset{
    mybox/.style={
        colback=white,
        colframe=black,
        sharp corners,
        boxrule=1pt,
        left=4mm,
        right=4mm,
        top=4mm,
        bottom=4mm,
        boxsep=2mm,        
        before skip=10pt plus 2pt minus 2pt, 
        after skip=10pt plus 2pt minus 2pt   
    }
}

\section{Additional Experiments}
\subsection{Effect of Backbone Language Models}
We analyze the effect of backbone language models by substituting the default GPT-2 with Llama3-8B, denoted by \textbf{KGTB (Llama3-8B)}. Llama3-8B has a larger size of parameters than GPT-2 and it is also the backbone of the most competitive baseline GNPR-SID. Tab.~\ref{tab:llama3} shows the performance comparison with GNPR-SID and the best baseline. Apparently, Llama3-8B has strengthened the leadership of KGTB, achieving an average 25.73\% gain against the best baseline. This observation suggests the stronger recommendation potential of larger language models. In addition, KGTB with much smaller GPT-2 still beats the best baseline, especially GNPR-SID built on Llama3-8B, further demonstrating the effectiveness of our proposed StruID and multi-behavior learning strategy.
\begin{table}[h!]
	\small
	\setlength{\tabcolsep}{1.5pt}
	\centering
	\begin{tabular}{c|c|cc|ccr}
		\toprule
		\textbf{Dataset} & \textbf{Metric}&\textbf{\makecell{GNPR\\-SID}}& \textbf{\makecell{Best\\Baseline}}& \textbf{KGTB} & \textbf{\makecell{KGTB\\(Llama3-8B)}}& \textbf{Imp.} \\
		\midrule
		\multirow{5}[0]{*}{\makecell[c]{FB}}
              &HR@1&0.1577&0.1577&\underline{0.1675}&\textbf{0.1993}&26.4\% \\
		     &HR@5&0.2827&0.2849&\underline{0.3190}&\textbf{0.3445}& 20.9\% \\
		   &N@5&0.2231&0.2231&\underline{0.2468}&\textbf{0.2762}& 23.8\% \\
		   &HR@10&0.3231&0.3494&\underline{0.3818}&\textbf{0.3946}& 12.9\% \\
              &N@10&0.2364&0.2366&\underline{0.2673}&\textbf{0.2924}& 23.6\% \\
		\midrule
		\multirow{5}[0]{*}{\makecell[c]{FP}}
              &HR@1&0.1649&0.1649&\underline{0.1716}&\textbf{0.1954}&18.5\% \\
		     &HR@5&0.2858&0.2858&\underline{0.3193}&\textbf{0.3321}&16.2\% \\
		   &N@5&0.2282&0.2282&\underline{0.2485}&\textbf{0.2677}& 17.3\% \\
		   &HR@10&0.3271&0.3440&\underline{0.3789}&\textbf{0.3802}& 10.5\% \\
              &N@10&0.2417&0.2417&\underline{0.2679}&\textbf{0.2833}& 17.2\% \\
		\midrule
		\multirow{5}[0]{*}{\makecell[c]{GA}}
              &HR@1 &0.1005&0.1005&\underline{0.1085}&\textbf{0.1429}&42.2\% \\
		     &HR@5 &0.1890&0.1982&\underline{0.2250}&\textbf{0.2597}& 31.0\% \\
		   &N@5  &0.1466&0.1476&\underline{0.1693}&\textbf{0.2043}& 38.4\% \\
		   &HR@10&0.2244&0.2506&\underline{0.2744}&\textbf{0.3106}& 23.9\% \\
              &N@10 &0.1582&0.1582&\underline{0.1854}&\textbf{0.2208}& 39.6\% \\
		\midrule
		\multirow{5}[0]{*}{\makecell[c]{GS}}
              &HR@1 &0.0911&0.0911&\underline{0.1065}&\textbf{0.1329}&45.9\% \\
		     &HR@5 &0.1736&0.1971&\underline{0.2258}&\textbf{0.2469}& 25.3\% \\
		   &N@5  &0.1350&0.1480&\underline{0.1696}&\textbf{0.1929}& 30.3\% \\
		   &HR@10&0.1999&0.2413&\underline{0.2706}&\textbf{0.2941}& 21.9\% \\
              &N@10 &0.1435&0.1618&\underline{0.1842}&\textbf{0.2082}& 28.7\% \\
		\bottomrule
	\end{tabular}%
	
    \caption{Performance comparison with GNPR-SID and the best baseline on four city datasets. }
    \label{tab:llama3}
\end{table}%

\subsection{Analysis of Auxiliary Behavior Learning}
\begin{figure}[h]
    \centering
    \includegraphics[width=0.38\textwidth]{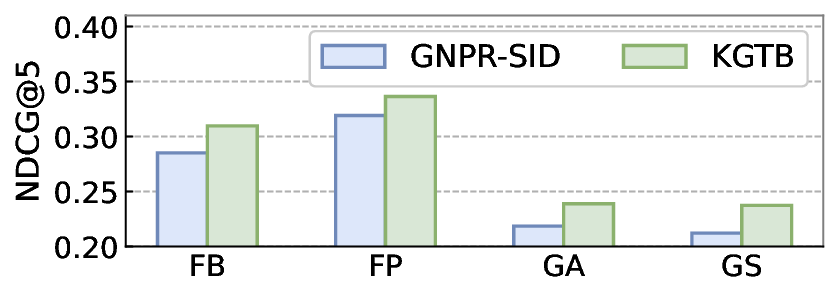}
    \caption{Performance comparison in predicting category visit behavior.}
    \label{fig:category}
\end{figure}

\begin{figure}[h]
    \centering
    \includegraphics[width=0.38\textwidth]{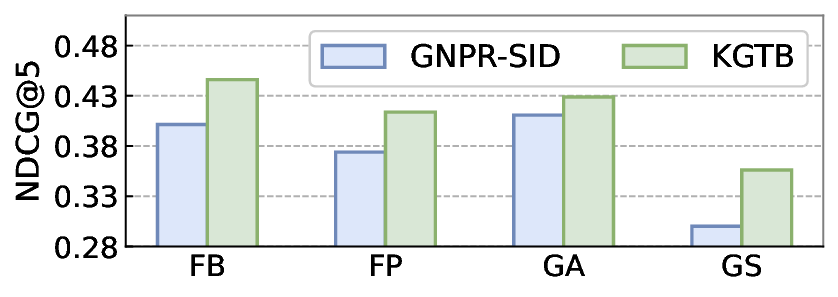}
    \caption{Performance comparison in predicting region visit behavior.}
    \label{fig:region}
\end{figure}

KGTB learns two auxiliary user behaviors, i.e., category and region visit behaviors, to enhance the LLM's mobility understanding. This experiment investigates how does KGTB perform in forecasting these two behaviors. As shown in Fig.~\ref{fig:category} and Fig.~\ref{fig:region}, KGTB outperforms the most competitive baseline GNPR-SID on both prediction tasks across four datasets. The findings indicates that KGTB has integrated knowledge of category and region visit behaviors, providing comprehensive insights into users' POI visit behavior.

\subsection{Analysis of Sequential Pattern Modeling}
The information loss issue in existing SIDs hinders LLMs to capture the sequential patterns between users' historical POIs and the target POIs. Such problem becomes even more pronounced when the target POIs have never been visited by the user, i.e., unseen POIs. In contrast, StruID provides rich heterogeneous information for sequential pattern modeling. To illustrate this, we investigate the performance of KGTB and GNPR-SID in recommending unseen POIs, and list the results in Tab.~\ref{tab:unseen}. 

As shown in Tab.~\ref{tab:unseen}, it is clear that the scores produced by both KGTB and GNPR are quite low, indicating that recommending unseen POIs is a very challenging task. Nevertheless, our proposed KGTB achieves an impressive improvement of 170.5\% on N@5 against GNPR-SID. This finding verifies the rich information delivered by StruID, which facilitates the modeling of sequential patterns.

\begin{table}[h]
	\centering
	\begin{tabular}{ccccc}
		\toprule
		Model & FB & FP & GA &GS\\
		\midrule
            GNPR-SID &0.0174&0.0124&0.0155&0.0094\\
		KGTB  &0.0307 &0.0280&0.0424&0.0382\\
        \midrule
	\end{tabular}%
        \caption{N@5 performance comparison on unseen POIs.}
	\label{tab:unseen}%
\end{table}%

%% file: main.bbl
\begin{thebibliography}{26}
\providecommand{\natexlab}[1]{#1}

\bibitem[{Chen et~al.(2025)Chen, Tao, Jiang, Liu, Yu, and
  Cong}]{chen2025enhancing}
Chen, Y.; Tao, Y.; Jiang, Y.; Liu, S.; Yu, H.; and Cong, G. 2025.
\newblock Enhancing Large Language Models for Mobility Analytics with Semantic
  Location Tokenization.
\newblock \emph{arXiv preprint arXiv:2506.11109}.

\bibitem[{Cheng et~al.(2013)Cheng, Yang, Lyu, and King}]{cheng2013you}
Cheng, C.; Yang, H.; Lyu, M.~R.; and King, I. 2013.
\newblock Where you like to go next: Successive point-of-interest
  recommendation.
\newblock In \emph{Proceedings of the 23th International Joint Conference on
  Artificial Intelligence}.

\bibitem[{Cho, Myers, and Leskovec(2011)}]{cho2011friendship}
Cho, E.; Myers, S.~A.; and Leskovec, J. 2011.
\newblock Friendship and mobility: user movement in location-based social
  networks.
\newblock In \emph{Proceedings of the 17th ACM SIGKDD international conference
  on Knowledge discovery and data mining}, 1082--1090.

\bibitem[{Feng et~al.(2015)Feng, Li, Zeng, Cong, and
  Chee}]{feng2015personalized}
Feng, S.; Li, X.; Zeng, Y.; Cong, G.; and Chee, Y.~M. 2015.
\newblock Personalized ranking metric embedding for next new poi
  recommendation.
\newblock In \emph{Proceedings of the 24th International Conference on
  Artificial Intelligence}, 2069--2075.

\bibitem[{Feng et~al.(2024)Feng, Lyu, Li, Sun, and Chen}]{feng2024move}
Feng, S.; Lyu, H.; Li, F.; Sun, Z.; and Chen, C. 2024.
\newblock Where to move next: Zero-shot generalization of llms for next poi
  recommendation.
\newblock In \emph{2024 IEEE Conference on Artificial Intelligence},
  1530--1535. IEEE.

\bibitem[{Geng et~al.(2022)Geng, Liu, Fu, Ge, and
  Zhang}]{geng2022recommendation}
Geng, S.; Liu, S.; Fu, Z.; Ge, Y.; and Zhang, Y. 2022.
\newblock Recommendation as language processing (rlp): A unified pretrain,
  personalized prompt \& predict paradigm (p5).
\newblock In \emph{Proceedings of the 16th ACM conference on recommender
  systems}, 299--315.

\bibitem[{Han et~al.(2025)Han, Yin, Chen, Jiang, Jiang, Li, Ma, Huang, Li, Jing
  et~al.}]{han2025mtgr}
Han, R.; Yin, B.; Chen, S.; Jiang, H.; Jiang, F.; Li, X.; Ma, C.; Huang, M.;
  Li, X.; Jing, C.; et~al. 2025.
\newblock MTGR: Industrial-Scale Generative Recommendation Framework in
  Meituan.
\newblock \emph{arXiv preprint arXiv:2505.18654}.

\bibitem[{Hu et~al.(2022)Hu, Shen, Wallis, Allen-Zhu, Li, Wang, Wang, Chen
  et~al.}]{hu2022lora}
Hu, E.~J.; Shen, Y.; Wallis, P.; Allen-Zhu, Z.; Li, Y.; Wang, S.; Wang, L.;
  Chen, W.; et~al. 2022.
\newblock Lora: Low-rank adaptation of large language models.
\newblock In \emph{The Tenth International Conference on Learning
  Representations}, volume~1, 3.

\bibitem[{Kong and Wu(2018)}]{kong2018hst}
Kong, D.; and Wu, F. 2018.
\newblock HST-LSTM: A hierarchical spatial-temporal long-short term memory
  network for location prediction.
\newblock In \emph{Proceedings of the 27th International Joint Conference on
  Artificial Intelligence}, 2341--2347.

\bibitem[{Lai et~al.(2024)Lai, Su, Wei, He, Wang, Chen, Zha, Liu, and
  Wang}]{lai2024disentangled}
Lai, Y.; Su, Y.; Wei, L.; He, T.; Wang, H.; Chen, G.; Zha, D.; Liu, Q.; and
  Wang, X. 2024.
\newblock Disentangled contrastive hypergraph learning for next POI
  recommendation.
\newblock In \emph{Proceedings of the 47th International ACM SIGIR Conference
  on Research and Development in Information Retrieval}, 1452--1462.

\bibitem[{Li et~al.(2024)Li, de~Rijke, Xue, Ao, Song, and Salim}]{li2024large}
Li, P.; de~Rijke, M.; Xue, H.; Ao, S.; Song, Y.; and Salim, F.~D. 2024.
\newblock Large language models for next point-of-interest recommendation.
\newblock In \emph{Proceedings of the 47th International ACM SIGIR Conference
  on Research and Development in Information Retrieval}, 1463--1472.

\bibitem[{Lian et~al.(2020)Lian, Wu, Ge, Xie, and Chen}]{lian2020geography}
Lian, D.; Wu, Y.; Ge, Y.; Xie, X.; and Chen, E. 2020.
\newblock Geography-aware sequential location recommendation.
\newblock In \emph{Proceedings of the 26th ACM SIGKDD international conference
  on knowledge discovery \& data mining}, 2009--2019.

\bibitem[{Lim et~al.(2022)Lim, Hooi, Ng, Goh, Weng, and
  Tan}]{lim2022hierarchical}
Lim, N.; Hooi, B.; Ng, S.-K.; Goh, Y.~L.; Weng, R.; and Tan, R. 2022.
\newblock Hierarchical multi-task graph recurrent network for next poi
  recommendation.
\newblock In \emph{Proceedings of the 45th international ACM SIGIR conference
  on Research and development in Information Retrieval}.

\bibitem[{Liu et~al.(2016)Liu, Wu, Wang, and Tan}]{liu2016predicting}
Liu, Q.; Wu, S.; Wang, L.; and Tan, T. 2016.
\newblock Predicting the next location: A recurrent model with spatial and
  temporal contexts.
\newblock In \emph{Proceedings of the 30th AAAI Conference on Artificial
  Intelligence}, 194--200.

\bibitem[{Rao et~al.(2022)Rao, Chen, Liu, Shang, Yao, and Han}]{rao2022graph}
Rao, X.; Chen, L.; Liu, Y.; Shang, S.; Yao, B.; and Han, P. 2022.
\newblock Graph-flashback network for next location recommendation.
\newblock In \emph{Proceedings of the 28th ACM SIGKDD Conference on Knowledge
  Discovery and Data Mining}, 1463--1471.

\bibitem[{Rao et~al.(2024)Rao, Jiang, Shang, Chen, Han, Yao, and
  Kalnis}]{rao2024next}
Rao, X.; Jiang, R.; Shang, S.; Chen, L.; Han, P.; Yao, B.; and Kalnis, P. 2024.
\newblock Next Point-of-Interest Recommendation with Adaptive Graph Contrastive
  Learning.
\newblock \emph{IEEE Transactions on Knowledge and Data Engineering}, 0(0):
  1--14.

\bibitem[{Schlichtkrull et~al.(2018)Schlichtkrull, Kipf, Bloem, Van Den~Berg,
  Titov, and Welling}]{schlichtkrull2018modeling}
Schlichtkrull, M.; Kipf, T.~N.; Bloem, P.; Van Den~Berg, R.; Titov, I.; and
  Welling, M. 2018.
\newblock Modeling relational data with graph convolutional networks.
\newblock In \emph{European semantic web conference}, 593--607. Springer.

\bibitem[{Wang et~al.(2025)Wang, Huang, Gao, Wang, Huang, and
  Shang}]{wang2025generative}
Wang, D.; Huang, Y.; Gao, S.; Wang, Y.; Huang, C.; and Shang, S. 2025.
\newblock Generative Next POI Recommendation with Semantic ID.
\newblock \emph{arXiv preprint arXiv:2506.01375}.

\bibitem[{Wang et~al.(2023{\natexlab{a}})Wang, Fang, Zeng, and
  Cheng}]{wang2023would}
Wang, X.; Fang, M.; Zeng, Z.; and Cheng, T. 2023{\natexlab{a}}.
\newblock Where would i go next? large language models as human mobility
  predictors.
\newblock \emph{arXiv preprint arXiv:2308.15197}.

\bibitem[{Wang et~al.(2023{\natexlab{b}})Wang, Zhu, Wang, Ma, Li, and
  Yu}]{wang2023adaptive}
Wang, Z.; Zhu, Y.; Wang, C.; Ma, W.; Li, B.; and Yu, J. 2023{\natexlab{b}}.
\newblock Adaptive Graph Representation Learning for Next POI Recommendation.
\newblock In \emph{Proceedings of the 46th International ACM SIGIR Conference
  on Research and Development in Information Retrieval}, 393--402.

\bibitem[{Wongso, Xue, and Salim(2024)}]{wongso2024genup}
Wongso, W.; Xue, H.; and Salim, F.~D. 2024.
\newblock GenUP: Generative User Profilers as In-Context Learners for Next POI
  Recommender Systems.
\newblock \emph{arXiv preprint arXiv:2410.20643}.

\bibitem[{Yang, Zhang, and Qu(2016)}]{yang2016participatory}
Yang, D.; Zhang, D.; and Qu, B. 2016.
\newblock Participatory cultural mapping based on collective behavior data in
  location-based social networks.
\newblock \emph{ACM Transactions on Intelligent Systems and Technology (TIST)},
  7(3): 1--23.

\bibitem[{Yang, Liu, and Zhao(2022)}]{yang2022getnext}
Yang, S.; Liu, J.; and Zhao, K. 2022.
\newblock GETNext: trajectory flow map enhanced transformer for next POI
  recommendation.
\newblock In \emph{Proceedings of the 45th International ACM SIGIR Conference
  on research and development in information retrieval}, 1144--1153.

\bibitem[{Yu et~al.(2020)Yu, Cui, Guo, Lu, Li, and Lu}]{yu2020category}
Yu, F.; Cui, L.; Guo, W.; Lu, X.; Li, Q.; and Lu, H. 2020.
\newblock A category-aware deep model for successive POI recommendation on
  sparse check-in data.
\newblock In \emph{Proceedings of the web conference 2020}, 1264--1274.

\bibitem[{Zhang et~al.(2018)Zhang, Tay, Yao, and Sun}]{zhang2018next}
Zhang, S.; Tay, Y.; Yao, L.; and Sun, A. 2018.
\newblock Next item recommendation with self-attention.
\newblock \emph{arXiv preprint arXiv:1808.06414}.

\bibitem[{Zhu et~al.(2024)Zhu, Jin, Liu, Qiu, Dong, and Li}]{zhu2024cost}
Zhu, J.; Jin, M.; Liu, Q.; Qiu, Z.; Dong, Z.; and Li, X. 2024.
\newblock CoST: Contrastive Quantization based Semantic Tokenization for
  Generative Recommendation.
\newblock In \emph{Proceedings of the 18th ACM Conference on Recommender
  Systems}, 969--974.

\end{thebibliography}
